\documentclass[prl,twocolumn,superscriptaddress,showpacs,amsmath,amssymb,aps]{revtex4-2}
\usepackage{amsfonts}
\usepackage{bm}
\usepackage{verbatim}

\usepackage{graphicx}

\begin{document}

\title{Suppressing the Kibble-Zurek mechanism by a symmetry-violating bias}

\author{ J.~Rysti}
\affiliation{Department of Applied Physics, Aalto University, POB 15100, FI-00076 AALTO, Finland}

\author{J.T.~M{\"a}kinen}
\email[]{jere.makinen@aalto.fi}
\affiliation{Department of Applied Physics, Aalto University, POB 15100, FI-00076 AALTO, Finland}

\author{S.~Autti}
\affiliation{Department of Applied Physics, Aalto University, POB 15100, FI-00076 AALTO, Finland}
\affiliation{Department of Physics, Lancaster University, Lancaster LA1 4YB, UK.}

\author{T.~Kamppinen}
\affiliation{Department of Applied Physics, Aalto University, POB 15100, FI-00076 AALTO, Finland}

\author{G.E.~Volovik}
\affiliation{Department of Applied Physics, Aalto University, POB 15100, FI-00076 AALTO, Finland}
\affiliation{L.D. Landau Institute for Theoretical Physics, Moscow, Russia}

\author{V.B.~Eltsov}
\affiliation{Department of Applied Physics, Aalto University, POB 15100, FI-00076 AALTO, Finland}

\date{\today}

\begin{abstract}

The formation of topological defects in continuous phase transitions is driven by the Kibble-Zurek mechanism. Here we study the formation of single- and half-quantum vortices during transition to the polar phase of $^3$He in the presence of a symmetry-breaking bias provided by the applied magnetic field. We find that vortex formation is suppressed exponentially when the length scale associated with the bias field becomes smaller than the Kibble-Zurek length. We thus demontrate an experimentally feasible shortcut to adiabaticity -- an important aspect for further understanding of phase transitions as well as for engineering applications such as quantum computers or simulators.

\end{abstract}

\maketitle

In continuous phase transitions, random local choice of the symmetry-breaking order parameter value leads to the formation of topological defects, such as quantized vortices. Originally a speculation in high-energy physics and cosmology \cite{Kibble1976}, this mechanism is now known as the Kibble-Zurek mechanism (KZM) \cite{Zurek1985,KZ_nature,KZ_nature2,Dziarmaga2005,Prufer2018}, and it is a cornerstone of out-of-equilibrium condensed matter physics. KZM has been observed in a wide range of systems such as superfluids, superconductors, and Bose condensates \cite{RevModPhys.83.863,DelCampo2014}. In the KZM scenario the transition takes place independently in various locations, and the characteristic size of these regions depends on the rate at which the transition is crossed. Each region inherits a random realization of the broken-symmetry feature of the new phase, such as the phase of the order parameter in a superfluid transition. When the expanding regions of the broken-symmetry phase merge, topological defects such as quantized vortices are formed. The predicted power-law dependence of the defect density on the quench rate has been confirmed in superfluid helium \cite{HQVs_prl,PhysRevB.90.024508} as well as other systems (see e.g. reviews \cite{0953-4075-50-2-022002,RevModPhys.83.863,DelCampo2014}). 

In the theory of broken-symmetry phase transitions, a symmetry-violating bias field plays an important role, initiating the choice between the different degenerate states beyond the critical point \cite{Sinai}. Bias can in particular be applied to non-adiabatic thermodynamic \cite{Kibble1976,Zurek1985} or quantum \cite{Dziarmaga2005} phase transitions that result in the formation of topological defects via the KZM. It has been proposed that if the applied symmetry-breaking bias is sufficiently large, the adiabatic (defect-free) regime is restored \cite{Dziarmaga2019}. The crossover from the non-adiabatic Kibble-Zurek regime to the adiabatic regime occurs at the characteristic value of the bias, which depends on the rate at which the critical point is crossed. Such crossover has been analyzed theoretically in a quantum phase transition in the Ising chain, see Refs.~\cite{Dziarmaga2019,PhysRevLett.126.070602} and in its classical counterpart in the Supplemental Material. Generally speaking, the KZM is expected to be modified in the presence of various external factors, such as inhomogeneities \cite{DelCampo2013}, or a propagating front of the phase transition \cite{KibbleVolovik1997}. Applying a bias, however, allows for the external control of the magnitude of the KZM directly. Controlled restoration of the adiabatic regime by a symmetry-breaking bias can be utilized in applications requiring delicate and fast control of engineered quantum systems \cite{Dziarmaga2019,del_Campo_2019}.

In this Letter we probe experimentally the use of an external bias for suppressing the formation of single quantum vortices (SQV) and half-quantum vortices (HQV)\cite{HQVs_prl,Makinen2019} produced by the Kibble-Zurek mechanism in the phase transition from normal $^3$He to the polar phase \cite{PolarDmitriev} of $^3$He. We report three central observations: (i) For HQVs the threshold bias for the onset of suppression is set by matching the characteristic length of the applied symmetry-breaking field to the Kibble-Zurek length set by the speed at which the transition is crossed. (ii) Beyond the onset, the suppression takes over exponentially, with the onset threshold normalizing the bias field in the exponent. (iii) The creation of SQVs is similarly suppressed for increasing bias fields while the threshold value is different from that for HQVs.


\begin{figure*}[tb!]
\centerline{\includegraphics[width=1 
\linewidth]{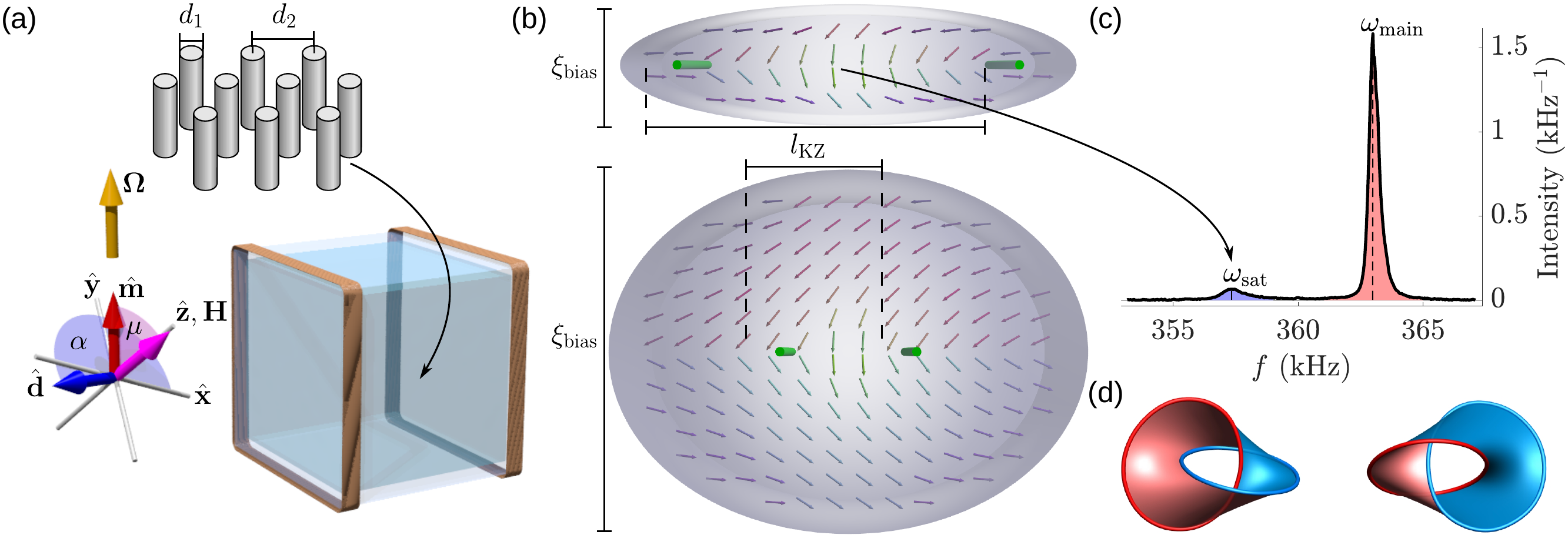}}
\caption{\label{experimental_setup} Experimental principles. (a) The cubic $4\times4\times4~\rm{mm}^3$ sample container is surrounded by rectangular NMR coils and is filled with solid strands oriented along the vertical axis with the average diameter $d_1 = 9\,$nm \cite{Asadchikov2015} and average separation $d_2 \approx 35\,$nm. The space between the strands is filled with liquid $^3$He. The magnetic field ${\mathbf{H}} \parallel \hat{\mathbf{z}}$ can be applied in any direction in the plane transverse to the NMR coil axes (angle $\mu$ is represented by the light red sector). The orbital anisotropy vector $\hat{\mathbf{m}}$ is pinned along the confining strands and the spin anisotropy vector $\hat{\mathbf{d}}$ is locked to the the $\mathbf{xy}$ plane (light blue sector represents angle $\alpha$) by the magnetic field. The sample can be rotated around the vertical axis with the angular velocity ${\bm \Omega}$ up to 3\,rad\,s$^{-1}$.
(b) The arrows represent the winding of $\hat{\mathbf{d}}$ in the vicinity of two HQV cores (green cylinders). On a loop around a HQV core the $\hat{\mathbf{d}}$ vector rotates by $\pi$. For a large applied bias (Top) pairs of HQVs are connected by narrow $\hat{\mathbf{d}}$-solitons (highlighted with the background color) and the width of the soliton, giving the characteristic length scale of the applied bias field $\xi_{\rm bias}$, is much smaller than the KZ length, $\xi_{\mathrm{bias}} \ll l_{\mathrm{KZ}}$, which results in suppression of the HQV formation in the phase transition to the superfluid phase. For a vanishing bias (Bottom) $\xi_{\mathrm{bias}} \gg l_{\mathrm{KZ}}$, the winding of the $\hat{\mathbf{d}}$-vector is nearly uniform, and formation of HQVs in the phase transition is not suppressed.
(c) The winding of the $\hat{\mathbf{d}}$ vector produces a trapping potential for spin waves, seen as a satellite peak in the NMR spectrum with the frequency $\omega_{\mathrm{sat}}$ which is smaller than the resonance frequency $\omega_{\mathrm{main}}$ for superfluid outside the solitons. The total area under the absorption spectrum is normalized to one.
(d) In the KZM topological defects form with random orientation and for example interconnected vortex rings shown here may form. The solitons connecting HQVs then form structures similar to Seifert surfaces \cite{Volovik2018}, shown here for two interconnected rings from two directions.}
\end{figure*}


The rich spectrum of topological defects in the polar phase, and the related bias fields one can apply, are understood in terms of the order parameter of the polar phase
\begin{equation}
A_{j \beta}=\Delta_{\rm P}{\bf \hat d}_j {\bf \hat m}_\beta e^{i\Phi} \,.
\label{OrderParameter}
\end{equation}
Here $\Delta_{\rm P}$ is the maximum gap in the quasiparticle energy spectrum and $\Phi$ is the superfluid phase. The unit vector ${\bf \hat d}$ determines the direction of the easy plane of the magnetic anisotropy and ${\bf \hat m}$ that of the orbital anisotropy. The anisotropy originates from p-wave Cooper pairing with the orbital momentum and spin of a pair equal to one. The polar phase is stabilized within the confining nanomaterial, which consists of nearly parallel solid strands, and ${\bf \hat m}$ is pinned along the strand direction, Fig.~\ref{experimental_setup}(a). The direction of ${\bf \hat d}$ is set by the competition between the magnetic anisotropy energy $\chi ({\bf \hat d} \cdot {\bf H})^2/2$ in the magnetic field ${\bf H}$, and the spin-orbit interaction energy $g_{\rm so} ({\bf \hat d} \cdot {\bf \hat m})^2$, where  $\chi$ is the magnetic susceptibility and $g_{\rm so}$ is the spin-orbit coupling. 

In large magnetic fields $H^2\gg H_{\rm so}^2=g_{\rm so} \chi^{-1}$, the ${\bf \hat d}$ vector is kept in the plane perpendicular to ${\bf H}$ and the spin-orbit interaction takes the form
\begin{equation}
F_{\rm so}=g_{\rm so}\sin^2\mu\sin^2\alpha\,,
\label{so2}
\end{equation}
where $\mu$ is the angle between the magnetic field and the ${\bf \hat m}$ vector and $\alpha$ is the azimuthal angle of ${\bf \hat d}$ in the plane perpendicular to the magnetic field. For $\mu\neq 0$, the spin-orbit interaction in Eq.~(\ref{so2})  lifts the degeneracy over $\alpha$, and thus it plays the role of the symmetry-violating bias with the magnitude $g_{\rm so}\sin^2\mu$. 

We first study entering the superfluid state from the normal state while applying a symmetry-breaking bias field via the spin-orbit coupling. For a magnetic field oriented along the strands, $\mu=0$, the spin-orbit bias is absent and the symmetry-breaking scheme (ignoring $SO(2)$ orbital rotations about $\hat{\bf m}$) is
\begin{equation}
G=U(1)\times SO(2) \rightarrow \Upsilon=Z_2\,.
\label{SymmetryBreaking}
\end{equation}
Here $G$ describes the symmetries of normal $^3$He, $U(1)$ is the symmetry under the phase transformation, and $SO(2)$ is the symmetry under rotation in spin space about the axis of the magnetic field. $\Upsilon$ denotes the symmetry of the polar phase order parameter, where $Z_2$ is the spin rotation by $\pi$ (corresponding to the change ${\bf \hat d} \rightarrow -{\bf \hat d}$) accompanied by the phase change by $\pi$. Since the homotopy group $\pi_1(G/\Upsilon)=Z\times Z \times Z_2$, this symmetry-breaking scheme leads to three types of topological defects: SQVs in the $\Phi$-field, spin vortices in the $\alpha$-field, and HQVs, where both $\Phi$ and $\alpha$ change by $\pi$.

When the spin-orbit interaction is turned on ($\mu\neq 0$), the $SO(2)$ symmetry in Eq.(\ref{SymmetryBreaking}) is explicitly violated, and one obtains the following symmetry-breaking scheme:
\begin{equation}
\tilde G=U(1) \rightarrow \tilde \Upsilon=1\,.
\label{SymmetryBreaking2}
\end{equation}
Now the homotopy group is $\pi_1(\tilde G/\tilde \Upsilon)=Z$, which means that only one of the three types of topological defects remain stable: the SQVs, which are not influenced by the spin-orbit interaction. Spin vortices and HQVs become termination lines of topological solitons \cite{MineyevVolovik1978,HQVs_prl,Makinen2019,PhysRevResearch.2.033013}, illustrated in Fig.~\ref{experimental_setup}(b). Assuming only HQVs are present, $\hat{\mathbf d}$-solitons connect pairs of HQVs of the opposite $\hat{\mathbf d}$-winding.

The presence of the solitons can be detected and their total volume in the sample measured using the nuclear magnetic resonance (NMR) techniques.The bulk of the sample forms the main peak in the continuous-wave NMR spectrum at the frequency $\omega_{\rm main}$. The $\hat{\mathbf{d}}$-soliton provides a trapping potential for standing spin waves, seen as a satellite peak in the NMR spectrum at the frequency $\omega_{\rm sat}$ \cite{HQVs_prl}. The relative sizes of the main peak and the satellite are determined by the volume occupied by the $\hat{\mathbf{d}}$-solitons in the sample. Measuring the initial density of KZ defects has traditionally been a complicated task due to the fast annihilation of non-equilibrium defects at temperatures close to the phase transition \cite{KZ_nature,KZ_nature2,PhysRevB.90.024508,ProgLowTempPhys_page9}. In our experiments the confining strands pin vortices in place once they are formed \cite{PhysRevResearch.2.033013,HQVs_prl,Makinen2019}, which provides the observer a frozen window to the out-of-equilibrium physics of the second-order phase transition and a direct measurement of the KZ vortex density.

We calibrate the size of the satellite peak by preparing a state by a very slow cooldown through the critical temperature $T_{\rm c}$ in zero magnetic field while the sample is in stable rotation. Under these conditions we create HQVs with aerial density given by $n_{\rm v} = 4\Omega \kappa^{-1}$, where $\Omega$ is the angular velocity, $\kappa=h/(2 m_3)$ is the quantum of circulation, $h$ is the Planck constant, and $m_3$ is the $^3$He atom mass. The calibration gives the relative satellite size $I_\mathrm{sat}=I_0 \sqrt{\Omega}$, where $I_0=0.090~\mathrm{s^{1/2}\,rad^{-1/2}}$ (see Supplemental Material). The inter-vortex distance assuming a square lattice is $L=n_{\rm v}^{-1/2}=\frac{1}{2} \sqrt{\kappa} I_0 I_\mathrm{sat}^{-1}$. We then use this relation to estimate the HQV density and intervortex distance for HQVs created purely by the KZM (i.e. for $\Omega = 0$). The combined effect of rotation and KZM is discussed in the Supplemental Material.

\begin{figure}
\centerline{\includegraphics[width=1\linewidth]{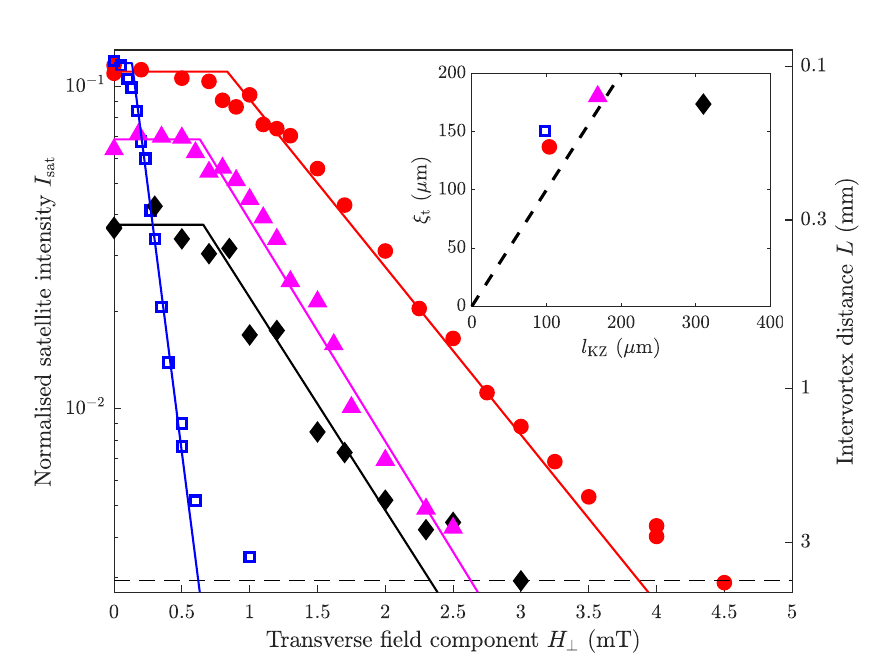}}
\caption{Suppression of the half-quantum vortex density created by the KZM as a function of the applied bias. Filled red circles, magenta triangles, and black diamonds correspond to quench rates of $\tau_{\rm Q} \approx 3.8\cdot10^{2}$~s, $\tau_{\rm Q} \approx 1.4 \cdot10^3$~s, and $\tau_{\rm Q} \approx 7.7\cdot10^3$~s, respectively, while applying a constant $H = 11$~mT magnetic field. The field is rotated to achieve different bias fields $H_{\perp} = H \sin \mu$. Open blue squares ($\tau_{\rm Q} \approx 6.0 \cdot 10^{2}$~s) correspond to measurements with zero axial field component, $H_\perp = H$. Vortex density is constant for $H_{\perp} < H_{\perp {\rm t}}$ and suppressed for higher bias fields. The suppression starts when the characteristic length scale of the bias field $\xi_{\mathrm{bias}}(H_\perp)$ becomes smaller than the Kibble-Zurek length $l_{\rm KZ}$. Solid lines correspond to theoretical model, see text for details. The dashed line shows where the intervotex distance becomes comparable with the container size. The inset shows the extracted bias length corresponding to the onset of the suppression of the KZM as a function of $l_{\rm KZ}$ with the same symbols. The dashed line is $\xi_{\rm t} = l_{\rm KZ}$.
\label{PolarPh_vortices}
} 
\end{figure}

We control the spin-orbit bias by applying a fixed magnetic field of $H=11$~mT with chosen transverse component $H_\perp=H\sin\mu$ during the cooldown through $T_{\rm c}$. We repeat cooldowns with different values of $H_\perp$ and different cooldown rates, Fig.~\ref{PolarPh_vortices}. We observe a constant satellite size for small $H_\perp$ and its gradual suppression for larger values of $H_\perp$. We suggest that the threshold field $H_{\perp\rm t}$ where the suppression of the formation of half-quantum vortices starts is determined by comparing the Kibble-Zurek length $l_{\rm KZ}= a \xi_0(\tau_{\rm Q}/\tau_0)^{1/4}$ with the characteristic length of the bias, $\xi_{\rm bias}$, given by the thickness of the $\hat{\bf d}$ solitons. Here $a \sim 1$ fixes the exact length scale for the defect formation (in our measurements $a \approx 2.3$, see Supplemental Material), the quench rate is $\tau_{\rm Q}^{-1}=-\mathrm{d}(T/T_{\rm c})/\mathrm{d}t|_{T=T_{\rm c}}$, $T$ is temperature, $t$ is time, $\xi_0$ is the superfluid coherence length at low temperature, $\tau_0 = \xi_0 v_{\mathrm{F}}^{-1} \sim 1\,$ns is the order parameter relaxation time, $v_{\mathrm{F}}$ is the Fermi velocity, $\xi_{\rm bias} \sim \xi_{\rm so}/\sin\mu$, and $\xi_{\rm so} = 17~\mu$m is the dipole length \cite{HQVs_prl}. 

Taking $L | _{H_{\perp}= 0}$ for transitions without bias as an estimate for $l_{\rm KZ}$ \cite{HQVs_prl, ProgLowTempPhys_page9, Bauerle1998}, and equating it with $\xi_{\rm bias}$ gives the following threshold bias for the suppression of HQV creation
\begin{equation}
H_{\perp {\rm t}} =  \frac{\xi_{\rm so}}{l_{\rm KZ}} H \,.
\label{CriterionH}
\end{equation}
In the spirit of Ref.~\cite{Dziarmaga2019} we propose that the defect density $\propto I_{\rm sat}^2$ decays exponentially after the transition field. In terms of the satellite intensity, this reads
\begin{equation}
    I_{\rm sat} =
\begin{cases}
  I_{\rm sat0} & \text{for } H_\perp < H_{\perp {\rm t}} \\
      I_{\rm sat0} \exp \left( 1-H_\perp/H_{\perp {\rm t}} \right) & \text{for } H_{\perp} \geq H_{\perp {\rm t}} \,,
\end{cases}
\label{eq:piecewise}
\end{equation}
where $I_{\rm sat0}$ is the initial satellite intensity. We note that for this model $\int_0^{\infty} I_{\rm sat} {\rm d}H_\perp = 2 I_{\rm sat0} H_{\perp {\rm t}}$ and the numerical integral of the measured $I_{\rm sat}$ can be used to determine $H_{\perp {\rm t}}$ without fitting.

Our experiments, Fig.~\ref{PolarPh_vortices}, confirm the validity of the model \eqref{eq:piecewise} up to the point when the inter-vortex distance becomes comparable to the sample size. We emphasize that the threshold field $H_{\perp {\rm t}}$ (which also normalizes the exponent) is determined by integration of the experimental data without fitting procedure. The result agrees well with the conjecture $\xi_{\rm t} \equiv \xi_{\rm bias}(H_{\perp {\rm t}}) = l_{\rm KZ}$.

The result for the slowest quench rate deviates, however, from this dependence. In the presence of a thermal gradient, the phase transitioncompared proceeds via a propagating front, where ordering of the low-temperature phase lags behind the temperature front where $T = T_{\rm c}$ by distance $l_{\rm F}$. The KZM operates in the band of width $l_{\rm F}$ and is generally modified in comparison to the homogeneous cooling scenario \cite{KibbleVolovik1997,PhysRevLett.83.116,PhysRevLett.125.260603}. As $\tau_{\rm Q}$ increases, $l_{\rm F}$ decreases and $l_{\rm KZ}$ increases. We suggest that the smaller of the two characteristic lengths, $l_{\rm KZ}$ and $l_{\rm F}$, determines the threshold bias $\xi_{\rm t}$. We estimate that in our measurement $l_{\rm F} < l_{\rm KZ}$ only for the slowest quench rate (black diamonds in Fig.~\ref{PolarPh_vortices}) for which $l_{\rm F} \sim 210\,\mu$m, matching the observed value at $\xi_{\rm t}$ (see Supplemental Materials for details).

Alternatively to the spin-orbit bias we can apply a direct field bias with a weak magnetic field oriented perpendicular to the orbital vector $\hat{\bf m}$. The small magnetic field $H_{\perp}<H_{\rm so}$ violates the symmetry under rotation about $\hat{\bf m}$, which leads to the formation of solitons, absent at zero magnetic field, with the soliton thickness now determined by the magnetic field directly, $\xi_{\rm bias}=\xi_{\rm H}=\xi_{\rm so} H_{\rm so}/H$. Equating $\xi_{\rm bias}$ with $l_{\mathrm{KZ}}$ yields a criterion for the threshold field similar to that in Eq.~\eqref{CriterionH} but with $H$ replaced by $H_{\rm so}$. The expected decrease of the threshold field in this case is $H_{\rm so}/H = \Omega_{\rm P}/\omega_{\rm main} \approx \sqrt{2 (1- \omega_{\rm sat} / \omega_{\rm main})}  \approx 0.17$ \cite{HQVs_prl}, where $\Omega_{\rm P}$ is the polar phase Leggett frequency. It is confirmed experimentally by the blue squares in Fig.~\ref{PolarPh_vortices}. Here the ratio of the threshold field relative to the red circles, which correspond to the spin-orbit bias with similar quench rate, is $\approx 0.16$.

\begin{figure}[tb]
\centerline{\includegraphics[width=1\linewidth]{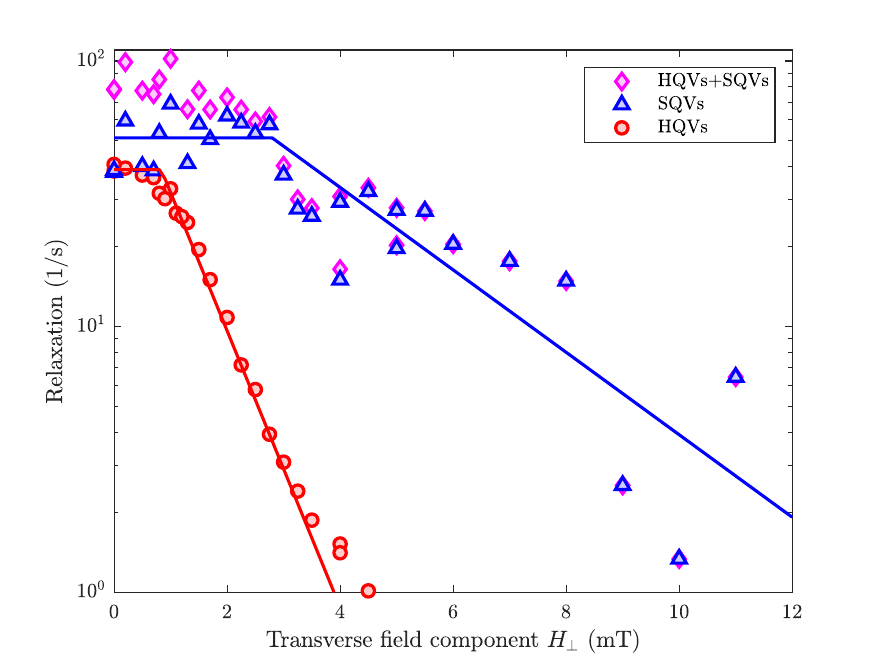}}
\caption{Suppression of SQV density as a function of the applied bias. The measured magnon BEC relaxation rate (magenta diamonds) include contributions from HQVs and SQVs. The HQV contribution (red circles) is separated using linear NMR measurements of the satellite intensity and calibration from Ref.~\cite{PhysRevResearch.2.033013}. The remaining contribution to relaxation (blue triangles) we attribute to SQVs. The observed relaxation rates are compared to the suppression model (solid lines), Eq.~\eqref{eq:piecewise}, and the red line corresponds to the same threshold as in Fig.~\ref{PolarPh_vortices}. In these measurements the quench rate is $\tau_{\rm Q}\sim 4\cdot 10^2$~s and the magnitude of the magnetic field is kept constant while its direction is varied. The largest transverse field value corresponds to $\mu=\pi/2$. Constant BEC relaxation rate not related to vortices has been subtracted.
\label{PolarPh_vortices_relaxation}
} 
\end{figure}

Finally, we study the fate of SQVs under the symmetry-breaking bias created by tilting the magnetic field. Due to the absence of the interconnecting topological solitons one would na\"{i}vely expect that such a bias has no effect on the KZM for SQVs. Without the solitons, SQVs are not seen in continuous-wave NMR, but they were found to increase the relaxation rate of a magnon BEC \cite{PhysRevLett.121.025303, PhysRevResearch.2.033013}. Independent measurements with SQVs created by rotation drive indicate that the relaxation rate increases monotonically when the SQV density grows \cite{PhysRevResearch.2.033013}. In our measurements we create both HQVs and SQVs by the KZM and subtract the effect of HQVs by using a calibration of the relaxation rate of the magnon BEC with respect to the satellite intensity, see Fig.~3b from Ref.~\cite{PhysRevResearch.2.033013}. The remaining contribution to the relaxation we attribute to SQVs. This contribution shows a characteristic dependence with a threshold and exponential suppression akin to Eq.~\eqref{eq:piecewise}, Fig.~\ref{PolarPh_vortices_relaxation}. A possible explanation for this behavior is that an applied bias influences the structure of the vortex core, which in $p$-wave superfluids can be complicated. For example in the bulk B phase, the $SO(2)$ symmetry of the SQV core is spontaneously broken at low temperatures and the core transforms into a pair of tightly-bound half-quantum cores \cite{Thuneberg1986,SalomaaVolovik1986,SalomaaVolovik1987}. In the polar phase, the spin-orbit interaction and magnetic anisotropy may play the role of the symmetry-violating bias for the phase transitions inside the vortex core, but the detailed investigation of this phenomenon remains a task for the future.

In conclusion, we report a crossover from the Kibble-Zurek regime of half-quantum vortex creation to the adiabatic regime, where vortex formation is rapidly suppressed by a symmetry-violating bias. We thus demonstrate an experimentally feasible shortcut to adiabaticity, where the adiabatic regime can be reached without an infinitely slow transition rate. In our experiments the symmetry-violating bias is provided either by the spin-orbit interaction or directly by external magnetic field. The crossover to the adiabatic regime takes place when the characteristic length scale of the bias, given by the thickness of the topological solitons connecting neighboring HQVs, becomes smaller than the KZ length determined by the transition rate or the thickness of the transition front in the case of slow inhomogeneous cooling. Beyond the onset, the suppression of the Kibble-Zurek mechanism takes place exponentially. We also report similar suppression of SQV formation by the KZM and provide preliminary evidence indicating that there may be a symmetry-breaking core transition in the SQV cores, sensitive to the symmetry-violating external bias.

The symmetry-breaking aspect of the bias field is essential for the suppression of the KZM, which otherwise is very robust. As an example, presented in the supplementary material, we show that adding an array of half-quantum vortices created by rotating the sample has no effect on the KZ mechanism even when the characteristic length scale of the added lattice becomes smaller than the KZ length. We also note that HQVs are composite defects, whose KZM formation is rarely studied experimentally, as well as analogs of Alice strings \cite{PhysRevD.103.054002,Chatterjee2017,Okada2015}. The KZM formation of HQVs studied here may shed light to defect formation across phase transitions in theories considering such systems.

Our results can be generalised to the bias-induced restoration of adiabadicity in various phase transitions including quantum phase transitions, which could provide applications in technologies such as quantum simulators and computers \cite{Dziarmaga2019,del_Campo_2019}. On a more speculative note, it is not excluded that bias plays a role in the so-called collapse of the wave function in quantum mechanics. In principle, the latter can be seen as ``phase transition'' occuring in the continuous spectrum of an infinite system \cite{Grady,Volovik2007,Ziaeepour2013}. One of the many quantum states participating in a given quantum superposition is perhaps then selected by the infinitesimal bias unavoidably present in any experiment.
 
\begin{acknowledgments}
We thank V.~V. Dmitriev for providing the NAFEN sample. This work has been supported by the European Research Council (ERC) under the European Union's Horizon 2020 research and innovation programme (Grant Agreement No. 694248) and by Academy of Finland project No. 332964. S.A. acknowledges support from the Jenny and Antti Wihuri Foundation via the Council of Finnish Foundations and T. K. acknowledges support from the Finnish Cultural Foundation. This research made use of the OtaNano – Low Temperature Laboratory infrastructure of Aalto University, that is part of the European Microkelvin Platform (European Union's Horizon 2020 Grant No. 824109).
\end{acknowledgments}
J.~R. and J.~T.~M. contributed equally to this work.



%

\renewcommand{\citenumfont}[1]{S#1}

\def\theequation{S\arabic{equation}}
\setcounter{equation}{0}

\def\thefigure{S\arabic{figure}}
\setcounter{figure}{0}

\clearpage

\section*{Supplemental Material}


\subsection*{Experimental setup}

We prepared the superfluid state by cooling the sample from above $T_{\rm c}$ to the polar phase using the ROTA nuclear demagnetization cryostat \cite{Heikkinen2014S}. The entire cryostat can be rotated up to 3~rad\,s$^{-1}$, and the axis of rotation is parallel to the confining strands, see Fig.~\ref{experimental_setup}(a). For confinement we used 94\% open NAFEN material (Al$_2$O$_3$, density 243~mg/cm$^3$) produced by AFN technology Ltd in Estonia. The Larmor frequency, defined by the magnetic field, was $f_{\rm L}=\omega_{\rm L}/2\pi=\gamma H/2\pi=363$~kHz, where $\gamma = 2 \pi \times 32.435$~MHz~T$^{-1}$ is the absolute value of the gyromagnetic ratio of $^3$He. The measurements were done at 7~bar pressure at which the critical temperature of the bulk superfluid is $T_{\rm c} \approx 1.6$~mK and the polar phase critical temperature within the aerogel sample was found to be $T_{\rm ca} \approx 0.972 T_{\rm c}$. Temperature was measured from the NMR spectrum of bulk B phase at high temperatures using the known B-phase Leggett frequency \cite{Thuneberg2001S,Hakonen1989S}, and from the NMR line in the polar phase in the axial (along NAFEN strands) field at low temperatures \cite{PolarDmitrievS,HQVs_prlS}. As a secondary thermometer we used a quartz tuning fork~\cite{2007_forksS, 2008_forksS, RiekkiForkS} located in bulk ${}^3$He. The container walls and NAFEN strands were preplated with a few atomic layers of ${}^4$He in order to order to stabilize the polar phase \cite{PhysRevLett.120.075301S} and to suppress the paramagnetic signal from solid ${}^3$He otherwise formed on all surfaces. The magnetic field could be freely rotated in the plane perpendicular to the NMR coils' axis from parallel to the strands ($\mu=0$) to transverse to them ($\mu=\pi/2$). Fast quenches through $T_{\rm ca}$ are done by heating the sample with a resistive heater above $T_{\rm c}$ and then switching the heater off. Slower transitions are done by controlling the temperature directly with the nuclear demagnetization stage. The quench rate $\tau_{\rm Q}$ is determined from the fork reading. During the transition magnetic field is applied at an angle and magnitude as described in the main text. To avoid formation of spin vortices, the NMR excitation is switched off during the cooldown. The NMR measurements are performed in transverse $(\mu = \pi/2)$ field at $T = 0.5\,T_{\rm c}$.

\subsection*{Ising chain under symmetry-breaking bias}

The restoration of adiabatic regime by a symmetry-violating bias was considered in Ref.~\cite{Dziarmaga2019S} building on toy model known as quantum Ising chain. Let us discuss the role of a symmetry-violating bias using its classical counterpart. The Hamiltonian of such a system, in terms of parameter $\beta$, may be written as
\begin{equation}
F= -\frac{1}{2}\cos^2\beta - g_\perp \sin\beta   - g_\parallel \cos\beta \,,
\label{Hamiltonian}
\end{equation}
where the term $g_\parallel \cos\beta$ plays the role of a symmetry-violating bias. For $g_\parallel = 0$ there is a discrete $Z_2$ symmetry $\cos\beta \rightarrow -\cos\beta $, and at $g_\perp=1$ there is a broken-symmetry phase transition between a paramagnetic and a ferromagnetic state. The minima of the free energy are located at
\begin{eqnarray}
\begin{aligned}
&\cos\beta=0 & \mathrm{for} \,\, g_\perp >1\,,\,\, g_\parallel=0 & \hspace{3mm} \mathrm{and}
\label{paramagnet}
\\
&\cos\beta=\pm \sqrt{1-g_\perp^2} & \mathrm{for}\,\, g_\perp <1\,,\,\, g_\parallel=0  & \hspace{3mm}
\label{ferromagnet}
\end{aligned}
\end{eqnarray}
for the paramagnetic and ferromagnetic phases, respectively. For $g_\parallel\neq 0$ the discrete $Z_2$ symmetry is explicitly broken, and the phase transition transforms into a smooth (adiabatic) crossover. This effect competes with the non-adiabatic Kibble-Zurek effect for the creation of topological defects: domain walls between two stable values of $\beta$. For each $\tau_{\rm Q}$ there exists a characteristic value of the symmetry-violating  bias, $g_{\parallel {\rm t}}$, which separates two regimes: the Kibble-Zurek regime of creation of domain walls at small bias $g_\parallel < g_{\parallel {\rm t}}$ and the adiabatic regime at large bias $g_\parallel > g_{\parallel {\rm t}}$, where the creation of the Kibble-Zurek domain walls begins to suppress.

\subsection*{Robustness of the KZ mechanism}



As an example of the robustness of the Kibble-Zurek mechanism (KZM), we consider the combined effect of the KZM and rotation drive which creates an additional density of the very same vortices. We observe that the KZ mechanism is robust against such direct external driving forces, a scenario touched theoretically in Ref.~\cite{PhysRevB.63.184501S}. That is, the vortex density created by the KZM is additive with the vortex density created by rotation.

In the KZM the average inter-vortex distance is given by the Kibble-Zurek length $l_{\rm KZ} = a \xi_0 (\tau_{\rm Q}/\tau_0)^{1/4}$, where $a \sim 1$ fixes the exact length scale for the defect formation, $\tau_{\rm Q}^{-1}=\left. -\frac{{\rm d}(T/T_{\rm c})}{{\rm d} t} \right|_{T=T_{\rm c}}$ is the cooldown rate at $T_{\rm c}$, $\xi_0 = \xi(T=0)$ is the coherence length at zero temperature, and the order-parameter relaxation time $\tau_0 = \xi_0 v_{\mathrm{F}}^{-1} \sim 1\,$ns, where $v_{\mathrm{F}}$ is the Fermi velocity. The vortices created by the KZM are randomly oriented, which in the case of half-quantum vortices (HQVs) complicates tracing the measured soliton volume back to vortex density as neighboring vortices are not aligned. In our case the vortex density is nevertheless very low, and the overwhelming majority of soliton volume therefore simply defined by the distance to the nearest vortex \cite{Volovik2018S}. In the low-density limit the relative intensity of the satellite signal becomes
\begin{equation}\label{KZ_law}
I_{\rm KZ} = \frac{1}{2} \frac{\xi_{\rm so}}{l_{\rm KZ}} \propto \tau_{\rm Q}^{-1/4}\,
\end{equation}
where $l_{\rm KZ}= a \xi_0(\tau_{\rm Q}/\tau_0)^{1/4}$.

HQVs can also be created by rotation drive. The resulting satellite peak intensity $I_{\rm sat}$ depends on the angular velocity $\Omega$ applied during the phase transition \cite{HQVs_prlS}:
 \begin{equation}
 I_{\Omega} = \Omega^{1/2} \xi_{\rm D} g_{\rm s} b/2 \equiv I_0 \Omega^{1/2}\,.
 \end{equation}
 Here $g_{\rm s} \sim 1.7$ is a numerical factor calculated in the supplemental material of Ref.~\cite{HQVs_prlS}, and $b \sim1$ describes the vortex lattice. In experiments this dependence is realized in the limit of vanishing cooldown rate.

 \begin{figure}[tb!]
\includegraphics[width=1 \linewidth]{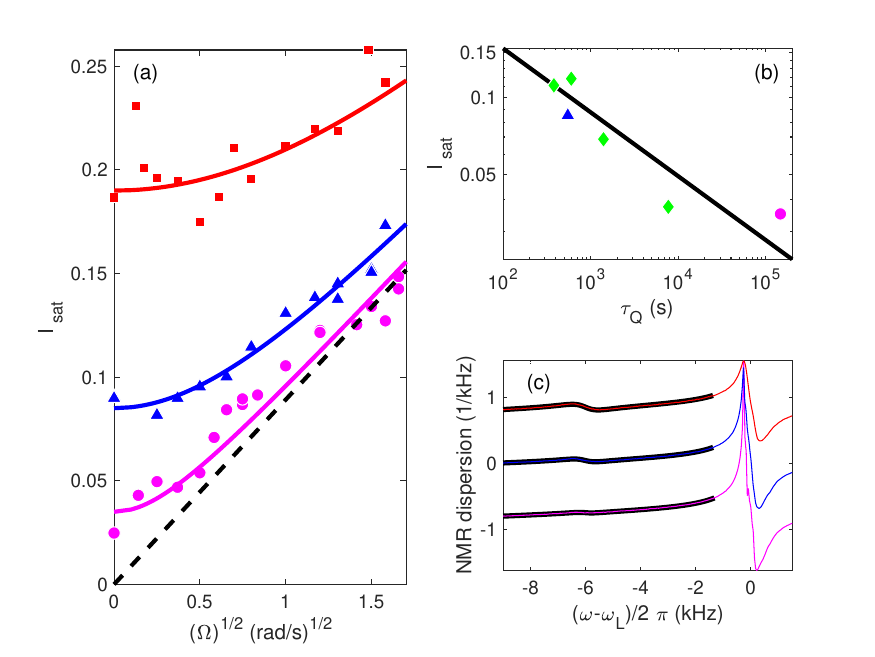}
\caption{\label{drivenKZM} 
{\bf Kibble-Zurek mechanism and rotation drive}: (a) The satellite intensity $I_{\rm sat}$ measured in slow ($\tau_{\rm Q} \approx 1.5 \cdot 10^ 5$~s, magenta circles) and fast ($\tau_{\rm Q} \approx 5.5\cdot 10^2$~s, blue triangles) zero-field cooldowns as a function of $\Omega$. The solid lines are fits that correspond to Eq.~(\ref{KZMbias}) with $I_0=0.090$~s\,rad$^{-1}$ and $I_{\rm KZ}$ fitted for each set separately. The dash line shows fitted equilibrium $I_{\rm sat}$, corresponding to $I_{\rm KZ}=0$. Applying rf drive at the resonance during the cooldown in the axial field creates spin vortices (red squares), seen as extra rotation-independent satellite intensity ($I_0 =0.090$~s\,rad$^{-1}$ here as well). Here $T=0.4T_{\rm c}$, and $P=6.9$~bar. (b) The satellite intensity $I_{\rm sat}$ measured in the absence of rotation and bias fields (green diamonds, $T=0.5T_{\rm c}$, $f_{\rm L}=363$~kHz, and $P=7.1$~bar, data from Fig.~2 in the main text) follows the Kibble-Zurek power law, Eq.~(\ref{KZ_law}), with $a = 2.3$ (solid line). The fitted satellite intensities at $\Omega = 0$ from panel (a) are marked with corresponding symbols. (c) The satellite intensities are extracted from the NMR dispersion (thin lines) spectrum by fitting a Lorentzian (thick black lines) to the satellite line. The spectra plotted here correspond to the zero-rotation cooldowns in panel (a). From bottom to top: slow cooldown, fast cooldown, and cooldown with rf pumping creating spin vortices. The spectra have been shifted vertically for clarity. All NMR spectra are recorded at $\mu=\pi/2$.
}
\end{figure}
 
In general, the ways the KZ vortex density and that created by the rotation drive combine can be expected to lie between two limiting scenarios: (a) The two mechanisms may be independent. In this case the vortex densities are additive and the satellite intensity becomes 

\begin{equation}
I_{\rm sat}= \sqrt{I_{\rm KZ}^2 + I_\Omega^2  } = \sqrt{I_{\rm KZ}^2 + I_0^2 \Omega  },
\label{KZMbias}
\end{equation}
where $I_{\rm KZ}$ and $I_\Omega$ stand for the satellite sizes created by either mechanism independently. (b) The rotation drive may merely act to polarize the KZ vortices until the KZ vortex density is exceeded by that required by the rotation applied. In this case the satellite intensity becomes $I_{\rm sat}=  {\rm max}[I_{\rm KZ} , I_\Omega ]$. 

The measured satellite intensity as a function of rotation velocity and cooldown rate is plotted in Fig.~\ref{drivenKZM}. There are two main observations: (i) The KZM is independent of applied rotation drive. Comparison of the measurements with the theoretical prediction for $I_{\rm sat}$ assuming rotation bias and KZM act in complete independence (Eq.~(\ref{KZMbias})) shows good quantitative agreement. The independence is emphasized especially by the fast-cooldown data for which $I_\Omega \approx I_{\rm KZ}$, therefore testing Eq.~(\ref{KZMbias}) in the most relevant range of parameters. The asymptotic dependence on $\Omega$ is confirmed by the data from slow cooldowns. The fitted value of $I_0$ is $I_0=0.090~\sqrt{\mathrm{s\,rad^{-1}}}$, while the theoretical prediction with $\xi_{\rm D} = 17~\mu$m is $I_0=0.11~\sqrt{\mathrm{s/rad}}$ \cite{HQVs_prlS}. The fitted averaged soliton length (vortex separation) of KZ vortices corresponds to $a = 2.3$, which is of expected order of magnitude \cite{Bauerle1998S,ProgLowTempPhys_page9S,HQVs_prlS}. (ii) Spin vortices can be created by applying a strong rf field on resonance during the phase transition. They are seen as additional satellite intensity. This additional vortex density does not change the effect of the rotation drive, which parallels the conclusion drawn above from observation (i). This confirms that the determination of vortex density from the soliton volume works as expected for independently formed topological objects. It should be noted that both these observations are likely to be valid only in the limit of low vortex density. At typical vortex densities in the experiments, the hard-core HQVs occupy no more than a few ppb of the sample volume, and even the soft-core spin vortices no more than a few ppm.

\subsection*{KZM in the wake of a thermal front}

In a realistic experimental scenario, the transition temperature is not crossed simultaneously in the whole sample volume. Any temperature gradient would result in a propagating thermal front, and as a result the symmetry-breaking phase transitions occurs in the wake of the propagating front\cite{PhysRevLett.83.116S, VolovikKibbleS, PhysRevLett.125.260603S}.

The volume containing the sample of the polar phase is open from one side to the heat exchanger volume and closed elsewhere, resulting in a thermal gradient across the sample. The temperature difference across the volume may be estimated by
\begin{equation}
\left| \frac{\Delta T}{\Delta x} \right| = \frac{\dot{Q}}{\mathcal{K} d^2}\,,
\end{equation}
where $\dot{Q}$ is the heat leak to the sample volume, $d = 4$~mm is the length of one side, and $\mathcal{K}$ is thermal conductivity. We estimate the total heat leak to the sample container to be $\sim 20$~pW from which we estimate the order of magnitude for the heat leak to the sample to be $\sim 1$~pW. Just above $T_{\rm c}$ the thermal conductivity of normal $^3$He at 7~bar pressure is $\mathcal{K} \sim 10^{4}$~erg\,(cm$\cdot$s$\cdot$K)$^{-1}$ \cite{PhysRevB.29.4933S, DobbsS}. We estimate the thermal conductivity in our sample in the direction perpendicular to NAFEN strands to be suppressed relative to the direction along the strands (assumed to be about that in bulk $^3$He) roughly by the same amount as spin diffusion is suppressed, i.e. by a factor of $\sim 8$ \cite{DmitrievAnisotropicSpinDiffS}. Under these assumptions an estimate of the temperature gradient across the sample is $\Delta T/\Delta x \sim 5 \times 10^{-6}$~mK\,mm$^{-1}$. The resulting thermal front propagation velocity in our experiments is in the range $v_{\rm T} \equiv (T_{\rm c} / \tau_{\rm Q} )/(\Delta T/\Delta x) \sim 40-900$~mm\,s$^{-1}$, where smaller values correspond to slower quenches.

The propagation velocity of the thermal front should be compared with the equilibration rate of the order parameter behind the front, which gives an estimate for the lowest front propagation velocity that still realizes the original KZM scenario \cite{VolovikKibbleS}
\begin{equation}
 v_{\rm Tc} \sim v_{\rm F} \left( \frac{\tau_0}{\tau_{\rm Q}} \right) ^{\frac{1}{4}}\,,
\end{equation}
where $v_{\rm F}$ is the Fermi velocity. For our experimental conditions, $v_{\rm Tc}$ lies in the range $30-60$~mm\,s$^{-1}$. We note that in this model, for the slowest quench rate in the experiments presented in the main text, $v_{\rm T} \sim v_{\rm Tc}$ and the initial vortex density is suppressed with respect to the original KZM scenario. This effect can be seen for solid black diamonds in Fig.~2 in the main text (same as the solid green diamond with the highest $\tau_{\rm Q}$ in Fig.~\ref{drivenKZM}).

We suggest that in addition to the lower initial vortex density, the slowly propagating front of the phase transition has another consequence. Unlike in the original KZ scenario, the vortices are now formed within a ``homogeneous'' region behind the front outside which the superfluid phase is already well defined. The thickness of this region becomes the new limiting length scale. According to Ref.~\cite{VolovikKibbleS} the thickness of this layer scales as $l_{\rm F} \approx v_{\rm T} \Delta t$, where
\begin{equation}
 \Delta t \approx \tau_{\rm Q} \frac{v_{\rm T}^2}{v_{\rm F}^2}\,.
\end{equation}
For the slowest quench rate in Fig.~\ref{PolarPh_vortices} (black diamonds), this gives $l_{\rm F} \approx 210\, \mu$m. We note that this is very close to the estimated $\xi_{\rm bias}(H_{\perp {\rm t}}) \approx 170 \, \mu$m while the KZ length extracted from the satellite intensity is $l_{\rm KZ} \approx 310\, \mu$m. We thus propose that the relevant length scale for the KZM for determination of the threshold bias is the smaller one of $l_{\rm KZ}$ and $l_{\rm F}$.


%

\clearpage


\end{document}